\documentclass[10pt,letterpaper]{IEEEtran}
\usepackage{amsmath}
\usepackage{graphicx}

\ifCLASSINFOpdf
\else
\fi

\begin{document}

\title{On the Anti-Jamming Performance of the NR-DCSK System}

\author{Binh Van Nguyen, Hyoyoung Jung, and Kiseon Kim
}

\maketitle

\begin{abstract}
This paper investigates the anti-jamming performance of the NR-DCSK system. We consider several practical jamming environments including broad-band jamming (BBJ), partial-time jamming (PTJ), tone jamming (TJ) consisting of both single-tone and multi-tone, and sweep jamming (SWJ). We first analytically derived the bit error rates of the considered system under the BBJ and the PTJ environments in closed-form expressions. Our derived results show that the system performances under these two jamming environments are enhanced as $P$ increases, where $P$ is the parameter of the NR-DCSK modulation scheme denoting the number of times a chaotic sample is repeated. In addition, our results demonstrate that for the PTJ, the optimal value of the jamming factor is close to zero when the jamming power is small, however, it increases and approaches one as the jamming power enlarges. We then investigate the performance of the considered system under the TJ and the SWJ environments via Monte-Carlo simulations. Our simulations show that single-tone jamming causes a more significant performance degradation than that provided by multi-tone jamming counterparts. Moreover, we point out that the system performance is significantly degraded when the starting frequency of the sweep jammer is close to the carrier frequency of the transmitted chaotic signals, the sweep bandwidth is small, and the sweep time is half of the transmitted bit duration.
\end{abstract}

\begin{IEEEkeywords}
Chaotic communication, noise reduction differential chaotic shift keying, jamming environments, bit-error-rate.
\end{IEEEkeywords}

\IEEEpeerreviewmaketitle

\section{Introduction}
\IEEEPARstart{C}{HAOTIC} communications (CC) has recently been considered as a promising alternative to the conventional direct sequence spread-spectrum (DS-SS) systems. The basic idea of CC is to replace pseudo-noise sequences by chaotic sequences, which can be directly generated by chaotic maps. Among various chaotic maps, logistic has been extensively exploited due to its simplicity and good performance \cite{Yang-14}. Chaotic systems are generally categorized as coherent and non-coherent chaotic systems. In the former systems, chaotic synchronization plays a vital role in deciding the systems performance. However, designing an efficient chaotic synchronization scheme still remains as a challenging problem. As a result, research works on the coherent chaotic systems are very limited. On the other hand, in the latter systems, data recovery procedure is much simpler than that of the coherent counterparts since the chaotic synchronization is not required. Consequently, a lot of efforts have been devoted to design effective non-coherent modulation schemes and to find novel applications of non-chaotic coherent systems in reality \cite{Kaddoum-16}.

The fundamental and most studied non-coherent modulation scheme is differential chaotic shift keying (DCSK). In a chaotic system with the DCSK scheme, referred to as DCSK system, a bit duration is divided into two equal time slots. The first slot is for transmitting a chaotic signal, namely as reference signal, which contains $\beta$ different chaotic samples. In addition, depending on the bit to be sent, the reference signal is either repeated or multiplied by a factor of $-1$ and transmitted in the second slot. At the receiver, the received signal is correlated with its "half a bit"-delayed version to estimate the transmitted information bit. The performances of the DCSK system over the additive white Gaussian noise (AWGN) and the multipath fading channels are extensively investigated in \cite{Sushchik-00}-\cite{Long-11} and \cite{Xia-04}, respectively. It is shown that enlarging the spreading factor beyond a certain value degrades the performance of the DCSK system. Although the DCSK system is simple and largely considered, it suffers from several drawbacks, i.e. low data rate, high energy consumption, using complex wideband delay, and so on. Motivated by this fact, several advanced systems have been recently proposed including reference modulated DCSK \cite{Yang-13}, high efficiency DCSK \cite{Yang-12}, improved DCSK \cite{Kaddoum-15}, short reference DCSK \cite{Kaddoum-16-2}, and noise reduction DCSK (NR-DCSK) \cite{Kaddoum-16-3}, among which the NR-DCSK system provides the best noisy performance. The generation of the reference signal in the NR-DCSK system is totally different from that in the DCSK counterpart. Particularly, only $\beta/P$ different chaotic samples are generated and each sample is repeated $P$ time to make a reference signal of $\beta$ samples in total.

Although chaotic systems in general and non-coherent chaotic systems in particular are partly designed for the anti-jamming (AJ) purpose, the AJ performance of a non-coherent chaotic system is only carried out in \cite{Lau-02}. Particularly, the performance of the DCSK system under the single-tone jamming (STJ) environment is studied. The authors derived the system bit-error-rate (BER) in a closed-form expression and show that the system performance is significantly degraded when the jamming frequency is an integer multiple of the bit frequency. Despite the fact that the fundamental DCSK system is already investigated, the AJ performances of its recent advanced variants have not been reported in the literature yet. To fill this gap, in this paper, we will consider the recent proposed NR-DCSK system and analyse its AJ performance under various practical jamming environments such as broad-band jamming (BBJ), tone jamming (TJ), partial-time jamming (PTJ), and sweep jamming (SWJ). We first analytically derived the BERs of the NR-DCKS system under the BBJ and PTJ in closed-form expressions. We then investigate the BER of the considered system under the TJ and the SWJ via Monte-Carlo simulations. Our analytical and simulation results reveal several novel insights about the effects of the jamming parameters on the system performance.

The remainder of this paper is organized as follows. Section II introduces the system model. Practical jamming signals are described in the section III. In section IV, we present mathematical frameworks to derive the system BERs under the BBJ and the PTJ environments. Thereafter, representative simulations are provided in section V to verify our analysis and further demonstrate the effects of the TJ and the SWJ on the performance of the considered system, followed by our conclusion in section VI.

\section{System Model}
The considered system consists of a pair of source-destination and a jammer, as illustrated in Fig. 1. Each node has a single antenna. In addition, the NR-DCSK scheme is employed for modulating and demodulating legitimate signals. Moreover, the logistic map is exploited for generating chaotic sequences. Here, the logistic map is considered because of its simplicity and good performance. A mathematical representation of the map is given by \cite{Yang-14}
\begin{align}\label{logistic}
  x_{k+1} = 1 - 2x_k^2,
\end{align}
where $E\left[ {{x_k}} \right] = 0$, $E\left[ {{x_k^2}} \right] = 1$, and $\text{var}\left[ {{x_k^2}} \right] = 1/2$ for normalized $x_k$.
\begin{figure}[!t]
  \includegraphics[width = 8cm]{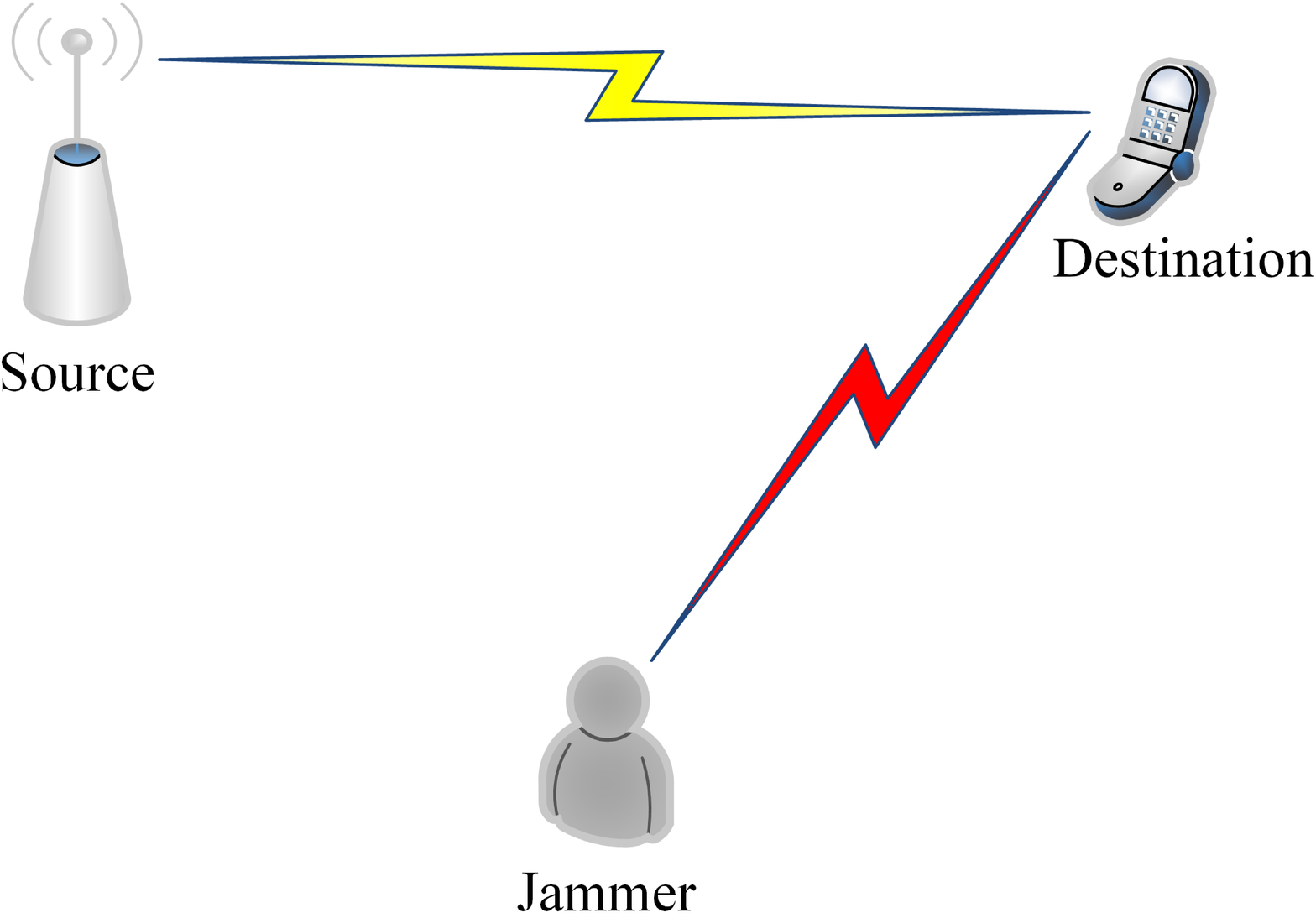}
  \caption{A NR-DCSK system in a jamming environment.}
\end{figure}

For the modulation process, each bit duration is divided into two equal time slots. The first time slot is allocated for transmitting a reference chaotic sequence length $\beta$, called spreading factor, while the second one is used for transmitting either the reference sequence (when the bit being sent is +1) or its inverted version (when the bit being sent is -1). The reference sequence is generated as follows. The chaotic generator first generates $\beta / P$ samples. Then, each chaotic sample is replicated $P$ times to obtain the reference sequence with length $\beta$ in total. As a result the transmitted signal of the $l^{th}$ bit, $b_l$, can be expressed as \cite{Kaddoum-16-3}
\begin{align}
{s_k^l} = \left\{ \begin{array}{l}
{x_{\left\lceil {\frac{k}{P}} \right\rceil}^ l}, \;\;\;\;\;\;\;\; \text{if} \;\; 0 < k \le \beta, \\
{b_l}{x_{\left\lceil {\frac{k}{P}} \right\rceil  - \beta }^l}, \; \text{if} \;\; \beta  < k \le 2\beta,
\end{array} \right.
\end{align}
where ${\left\lceil {\cdot} \right\rceil }$ denotes the ceiling operator. Graphically, $s_k^l$ can be illustrated as in Fig. 2.
\begin{figure}[!t]
  \includegraphics[width = 7cm]{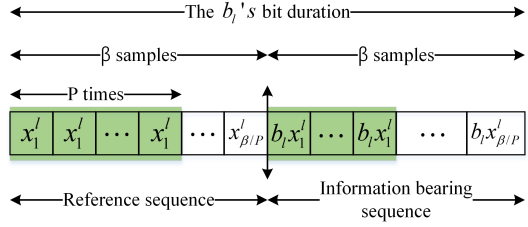}
  \caption{A graphical representation of the transmitted signal of the bit $b_l$.}
\end{figure}
\begin{figure}[!t]
  \includegraphics[width = 7cm]{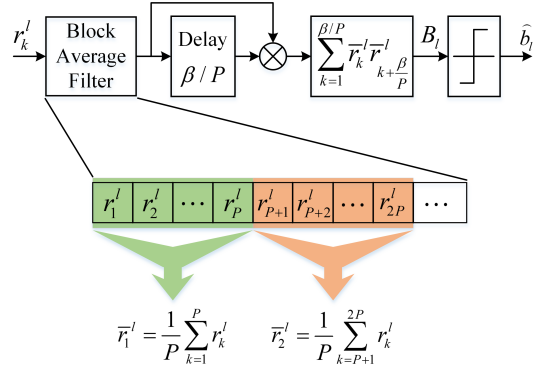}
  \caption{Receiver's structure and operation of the NR-DCSK system.}
\end{figure}

The received baseband signal at the receiver can be express as follows
\begin{align}
r_k^l = s_k^l + j_k^l + n_k^l,
\end{align}
where $j_k^l$ and $n_k^l$ are the jamming signal and the AWGN. Here we consider the AWGN channel so that we can solely focus on the effect of jamming on the system performance. The assumption of the AWGN channel has also been adopted in several related works, i.e. \cite{Lau-02}-\cite{Lau-04}. The received signal is first averaged using a block average filter (BAF) with a block size of $P$. The averaged signal is then correlated with its replica that is time delayed by a factor of $\beta / P$. Thereafter, the resultant correlated signal is summed over $\beta / P$ samples and passed through a threshold detector to recover the transmitted information bit. The BAF is used to reduce the power of the AWGN, from which the noisy performance enhancement of the NR-DCSK can be achieved. A detailed illustration of the receiver is given in Fig. 3, in which the output of the BAF would be given as
\begin{align}
\bar r_k^l &= \frac{1}{P}\sum\limits_{p = \left( {k - 1} \right)P + 1}^{kP} {r_p^l} \nonumber \\
& = s_k^l + \frac{1}{P}\sum\limits_{p = \left( {k - 1} \right)P + 1}^{kP} {j_p^l}  + \frac{1}{P}\sum\limits_{p = \left( {k - 1} \right)P + 1}^{kP} {n_p^l}.
\end{align}
In addition,  the decision variable at the input of the threshold detector can be written as
\begin{align}
& {B_l} = \sum\limits_{k = 1}^{\beta /P} {\left( {x_k^l + \frac{1}{P}\sum\limits_{p = \left( {k - 1} \right)P + 1}^{kP} {j_p^l}  + \frac{1}{P}\sum\limits_{p = \left( {k - 1} \right)P + 1}^{kP} {n_p^l} } \right)} \nonumber \\
& \cdot \left( {{b_l}x_k^l + \frac{1}{P}\sum\limits_{p = \left( {k - 1} \right)P + 1}^{kP} {j_{p + \beta}^l}  + \frac{1}{P}\sum\limits_{p = \left( {k - 1} \right)P + 1}^{kP} {n_{p + \beta}^l} } \right).
\end{align}

\section{Jamming Models}
In this section, we shall present several jamming types that are commonly encountered in military systems. These common jamming types include BBJ, PBJ, TJ, and SWJ.

\subsection{Broad-band Jamming \cite{Poisel-11}}
A broad-band jammer places jamming energy across the entire frequency bandwidth used by the target communication system. The BBJ essentially raises the background noise level at the receiver, creating a higher noise environment, which makes it more difficult for the target communication system to operate. At the very least it decreases the range over which the target communication system is effective. Since BBJ generates signals that are similar to broadband background noise, it is commonly modeled as a zero-mean Gaussian random variable with variance (power) $P_j$.

\subsection{Partial-time Jamming \cite{Xu-07}}
A partial time jammer turns on and off periodically according to the jamming factor $\rho = \frac{T_{on}}{T_{on} + T_{off}}$, where $T_{on}$ and $T_{off}$ denote the periods in which the jammer is on and off, respectively. In addition, the summation of $T_{on}$ and $T_{off}$ is called the jamming duty cycle. When the jammer is on, it is commonly assumed to emits broad-band signals, i.e. Gaussian noise, with power $P_j / \rho$, where $P_j$ is the average jamming power. It means that the jammer turns on and off periodically to focus more power on degrading the communication of the target system. It is noteworthy that the BBJ is a special case of the PTJ when $\rho = 1$.

\subsection{Tone Jamming \cite{Mao-17}}
In TJ, one ore more jammer tones are strategically placed in the target signals' spectrum. TJ includes STJ and multiple-tone jamming (MTJ). STJ places a single-tone where it is needed, while MTJ equally distributes the jamming power among several tones. Mathematically, the TJ can be expressed as follows
\begin{align}
{J_{tj}}\left( t \right) = \sum\limits_{m = 1}^M {\sqrt{\frac{2P_j}{M}}\sin \left( {2\pi {f_m}t + {\theta _m}} \right)},
\end{align}
where $M$, $P_j$, $f_m$, and $\theta_m$ are the number of jamming tones, total jamming power, and the frequency and the phase of the $m^{th}$ jamming tone, respectively. In addition, it should be noticed that $f_m$ here is the baseband frequency of the $m^{th}$ jamming tone, which is different from the passband counterpart, i.e. $f_m + f_c$ where $f_c$ denotes the carrier frequency of the transmitted chaotic signals.

\subsection{Sweep Jamming \cite{Chen-16}}
A sweep jammer rapidly sweeps a narrow band jamming signal over a wide frequency band. The SWJ can be characterized by its starting frequency $f_{start}$, stopping frequency $f_{stop}$, sweep rate $\Delta f$, and sweep time $T_{sw}$. In addition, there are several variants of the SWJ depending on narrow-band signals being used and sweep methods that specify the evolution of the instantaneous sweep frequency. Generally, narrow-band signals used to generate the SWJ include sinusoidal, triangular pulses, and rectangular pulses. Moreover, sweep methods contain linear, quadratic, and logarithmic. Among various aforementioned options, the one based on sinusoidal signals and linear sweep method is commonly used and modeled as follows
\begin{align}
{J_{sw}}\left( t \right) = \sqrt {2{P_j}} \sin \left( {2\pi {f_{start}}t + \pi \Delta f {t^2} + \theta_{sw} } \right),
\end{align}
where $\Delta f = \frac{f_{stop} - f_{start}}{T_{sw}}$ and $\theta_{sw}$ denotes the initial phase of the sweep jamming signals. Similar to the case of the TJ, $f_{start}$ and $f_{stop}$ here are all the baseband frequencies.

\section{Anti-Jamming Performance Analysis}
In this section, we analytically analyse the BER of the NR-DCSK system under the PTJ environment. In addition, since the BBJ is a special case of the PTJ, the system BER under the BBJ environment can be straightforwardly obtained from that under the PTJ. The system performances under the other types of jamming will be evaluated via Monte-Carlo numerical simulations in the next section due to mathematical intractability.

The decision variable presented in (5) can be further expanded as follows
\begin{align}
& {B_l} = \underbrace {\sum\limits_{k = 1}^{\beta /P} {{b_l}{{\left( {x_k^l} \right)}^2}} }_U + \underbrace {\sum\limits_{k = 1}^{\beta /P} {\frac{{x_k^l}}{P}} \sum\limits_{p = \left( {k - 1} \right)P + 1}^{kP} {j_{p + \beta }^l} }_{{I_1}} \nonumber \\
& + \underbrace {\sum\limits_{k = 1}^{\beta /P} {\frac{{{b_l}x_k^l}}{P}} \sum\limits_{p = \left( {k - 1} \right)P + 1}^{kP} {j_p^l} }_{{I_2}} \nonumber \\
& + \underbrace {\sum\limits_{k = 1}^{\beta /P} {\frac{1}{{{P^2}}}} \sum\limits_{p = \left( {k - 1} \right)P + 1}^{kP} {j_p^l} \sum\limits_{p = \left( {k - 1} \right)P + 1}^{kP} {j_{p + \beta }^l} }_{{I_3}} \nonumber \\
& + \underbrace {\sum\limits_{k = 1}^{\beta /P} {\frac{{x_k^l}}{P}} \sum\limits_{p = \left( {k - 1} \right)P + 1}^{kP} {n_{p + \beta }^l} }_{{N_1}} + \underbrace {\sum\limits_{k = 1}^{\beta /P} {\frac{{{b_l}x_k^l}}{P}} \sum\limits_{p = \left( {k - 1} \right)P + 1}^{kP} {n_p^l} }_{{N_2}} \nonumber \\
& + \underbrace {\sum\limits_{k = 1}^{\beta /P} {\frac{1}{{{P^2}}}} \sum\limits_{p = \left( {k - 1} \right)P + 1}^{kP} {j_p^l} \sum\limits_{p = \left( {k - 1} \right)P + 1}^{kP} {n_{p + \beta }^l} }_{{N_3}} \nonumber \\
& + \underbrace {\sum\limits_{k = 1}^{\beta /P} {\frac{1}{{{P^2}}}} \sum\limits_{p = \left( {k - 1} \right)P + 1}^{kP} {n_p^l} \sum\limits_{p = \left( {k - 1} \right)P + 1}^{kP} {j_{p + \beta }^l} }_{{N_4}} \nonumber
\end{align}
\begin{align}
+ \underbrace {\sum\limits_{k = 1}^{\beta /P} {\frac{1}{{{P^2}}}} \sum\limits_{p = \left( {k - 1} \right)P + 1}^{kP} {n_p^l} \sum\limits_{p = \left( {k - 1} \right)P + 1}^{kP} {n_{p + \beta }^l} }_{{N_5}},
\end{align}
where $U$, $I_i$, and $N_j$ represent the desired, interference, and noise components. It is noteworthy that $U$, $I_i$, and $N_j$ are independent because the chaotic signal, jamming signal, and AWGN are independent. In addition, at any given time, the chaotic signal is independent of its time-delayed version. Under the PTJ condition, the jamming signal $j_k$ is modeled as a zero-mean Gaussian random variable with variance (power) $P_j / \rho$, where $P_j$ is the average jamming power. In addition, the power of the AWGN $n_k$ is assumed to be $N_0/2$.

Following the central limit theorem, the decision variable $B_l$ can be assumed to follows the Gaussian distribution, and thus, the error probability of transmitting the bit $b_l$ can be calculated as follows
\begin{align}
BE{R_l} &= \frac{1}{2}\Pr \left[ {{B_l} < 0|{b_l} =  + 1} \right] + \frac{1}{2}\Pr \left[ {{B_l} > 0|{b_l} =  - 1} \right] \nonumber \\
& = \frac{1}{2}erfc\left[ {{{\left( {\frac{{2{\mathop{\rm var}} \left[ {{B_l}} \right]}}{{{{\left( {E\left[ {{B_l}} \right]} \right)}^2}}}} \right)}^{ - 1/2}}} \right],
\end{align}
where $erfc\left(  \cdot  \right)$, ${\mathop{\rm var}} \left[ {{\cdot}} \right]$, and $E\left[ {{\cdot}} \right]$ are the complementary error function, the variance, and the mean operations, respectively. Equation (9) indicates that to obtain $BER_l$, the mean and variance of $B_l$ are required. Firstly, the mean of $B_l$ can be readily derived as follows
\begin{align}
E\left[ {{B_l}} \right] &= E\left[ U \right] = {b_l}E\left[ {\sum\limits_{k = 1}^{\beta /P} {{{\left( {x_k^l} \right)}^2}} } \right] = {b_l}{E_P}.
\end{align}

To derive the variance of $B_l$, it is necessary to consider whether the jammer is on or off during the transmission time of the bit $b_l$. Let's first consider the case that the jammer is on. Under this condition, the bit $b_l$ can be assumed to be jammed with a probability of $\rho$. With a typical assumption of a high spreading factor, the bit energy ${E_b} = 2P\sum\limits_{k = 1}^{\beta /P} {E\left[ {{{\left( {x_k^l} \right)}^2}} \right]}  = 2P{E_P}$ can be considered to be constant \cite{Kaddoum-15}-\cite{Kaddoum-16-3}. In addition, we can readily show that the average of $P$ independent and identically distributed (i.i.d.) samples of a zero-mean Gaussian variable with variance of $X$ is still a zero-mean Gaussian random variable with a reduced variance of $X/P$. Based on this results, we can derive the following results
\begin{align}
{\rm var}\left[ {{B_l}} \right] = \sum\limits_{i = 1}^3 {{\mathop{\rm var}} \left[ {{I_i}} \right]}  + \sum\limits_{j = 1}^5 {{\mathop{\rm var}} \left[ {{N_j}} \right]},
\end{align}
where ${{\mathop{\rm var}} \left[ {{I_i}} \right]}$ and ${{\mathop{\rm var}} \left[ {{N_j}} \right]}$ are given by
\begin{align}
{\mathop{\rm var}} \left[ {{I_1}} \right] = \frac{{{E_P}{P_j}}}{\rho P} = {\mathop{\rm var}} \left[ {{I_2}} \right],
\end{align}
\begin{align}
{\mathop{\rm var}} \left[ {{I_3}} \right] &= \frac{{\beta P_j^2}}{{{\rho^2 P^3}}}, \\
{\mathop{\rm var}} \left[ {{N_1}} \right] &= \frac{{{E_P}{N_0}}}{{2P}} = {\mathop{\rm var}} \left[ {{N_2}} \right], \\
{\mathop{\rm var}} \left[ {{N_3}} \right] &= \frac{{\beta {P_j}{N_0}}}{{2 \rho {P^3}}} = {\mathop{\rm var}} \left[ {{N_4}} \right], \\
{\mathop{\rm var}} \left[ {{N_5}} \right] &= \frac{{\beta N_0^2}}{{4{P^3}}}.
\end{align}
Plugging all the results given in (10)-(16) into (9), yields the following error probability of transmitting the bit $b_l$ while the PTJ jammer is on
\begin{align}
BER_l^{on} &= \frac{1}{2}erfc\left[ {\left( {\frac{{4{P_j / \rho} + 2{N_0}}}{{P{E_P}}}} \right.} \right. \nonumber \\
& \;\;\;\;\;\;\;\;\;\;\;\;\;\;\; \left. {{{\left. { + \beta \frac{{2{P_j}{N_0} / \rho + 2P_j^2 / \rho^2 + N_0^2/2}}{{{P^3}E_P^2}}} \right)}^{ - 1/2}}} \right].
\end{align}
On the other hand, when the partial-time jammer is off during the transmission time of the bit $b_l$, we can readily obtain the corresponding error probability by setting $P_j = 0$ in (17). That is,
\begin{align}
BER_l^{off} = \frac{1}{2}erfc\left[ {{{\left( {\frac{{2{N_0}}}{{P{E_P}}} + \beta \frac{{N_0^2/2}}{{{P^3}E_P^2}}} \right)}^{ - 1/2}}} \right].
\end{align}
Then, a weighted summation of (17) and (18), whose weighting factors are the probabilities that the bit $b_l$ is jammed and un-jammed, produces
\begin{align}
BER_l &= \frac{\rho}{2}erfc\left[ {\left( {\frac{{4{P_j / \rho} + 2{N_0}}}{{P{E_P}}}} \right.} \right. \nonumber \\
& \;\;\;\;\;\;\;\;\;\;\;\;\;\;\; \left. {{{\left. { + \beta \frac{{2{P_j}{N_0} / \rho + 2P_j^2 / \rho^2 + N_0^2/2}}{{{P^3}E_P^2}}} \right)}^{ - 1/2}}} \right] \nonumber \\
& + \frac{1-\rho}{2}erfc\left[ {{{\left( {\frac{{2{N_0}}}{{P{E_P}}} + \beta \frac{{N_0^2/2}}{{{P^3}E_P^2}}} \right)}^{ - 1/2}}} \right].
\end{align}
We observe that the above $BER_l$ does not depend on the bit index $l$, and thus, the system BER under the PTJ environment can be expressed as follows
\begin{align}
BER_{ptj} &= \frac{\rho }{2}erfc\left[ {{{\left( {8\frac{{\frac{{{P_j}}}{\rho } + \frac{{{N_0}}}{2}}}{{{E_b}}} + 8\beta \frac{{{{\left( {\frac{{{P_j}}}{\rho } + \frac{{{N_0}}}{2}} \right)}^2}}}{{PE_b^2}}} \right)}^{ - 1/2}}} \right] \nonumber \\
& + \frac{{1 - \rho }}{2}erfc\left[ {{{\left( {\frac{{4{N_0}}}{{{E_b}}} + \frac{{2\beta N_0^2}}{{PE_b^2}}} \right)}^{ - 1/2}}} \right].
\end{align}

Under the special case of the BBJ environment, the system BER can also be readily obtained from (20) by setting $\rho = 1$. Particularly,
\begin{align}
BER_{bbj} = \frac{1}{2}erfc\left[ {\left( {8\frac{{{P_j} + \frac{{{N_0}}}{2}}}{{{E_b}}}} \right.} \right.\left. {{{\left. { + 8\beta \frac{{\left( {{P_j} + \frac{{{N_0}}}{2}} \right)}^2}{{PE_b^2}}} \right)}^{ - 1/2}}} \right].
\end{align}
It is observed from (20) and (21) that as $P$ increases, the BERs are reduced and tend to the lower bound given by
\begin{align}
BER_{ptj}^{lower} &= \frac{\rho }{2}erfc\left[ {{{\left( {\frac{8}{{{E_b}}}\left( {\frac{{{P_j}}}{\rho } + \frac{{{N_0}}}{2}} \right)} \right)}^{ - 1/2}}} \right] \nonumber \\
& + \frac{{1 - \rho }}{2}erfc\left[ {{{\left( {\frac{{4{N_0}}}{{{E_b}}}} \right)}^{ - 1/2}}} \right].
\end{align}
Taking a closer look at (20) and (22), we can observe that when $P_j$ is large enough, the partial-time jammer should select $\rho = 1$ to optimally degrade the performance of the target system. On the other hand, when $P_j$ is small, it is obvious that $\rho$ should be chosen close to zero to increase the instantaneous jamming power, and thus, further deteriorates the performance of the target system.

\section{Simulation Results}
In this section, we shall provide several representative simulation results to (i) verify our analysis on the system BERs under the BBJ and the PTJ environments and (ii) illustrate the effects of the TJ and the SWJ on the system performance.

In Fig. 4, we simulate the BER of the NR-DCSK system versus $\frac{E_b}{N_0}$ under the PTJ and the BBJ environments with $JSR = \frac{P_j}{P_s} = 5$ dB. The figure first shows that the analysis curves follow the simulated ones closely, which validates our analysis. Secondly, we observe that with $JSR = 5$ dB, the system BER is enhanced as the jamming factor $\rho$ increases. In other words, from the jammer's perspective, the jammer is better to be active with a short duration and transmit with a higher power to more severely degrade the target system performance. Lastly, the figure confirms that as $P$ enlarges, the system BER improves and tends to the lower bound given in (22).
\begin{figure}[!t]
  \includegraphics[width = 8.5cm]{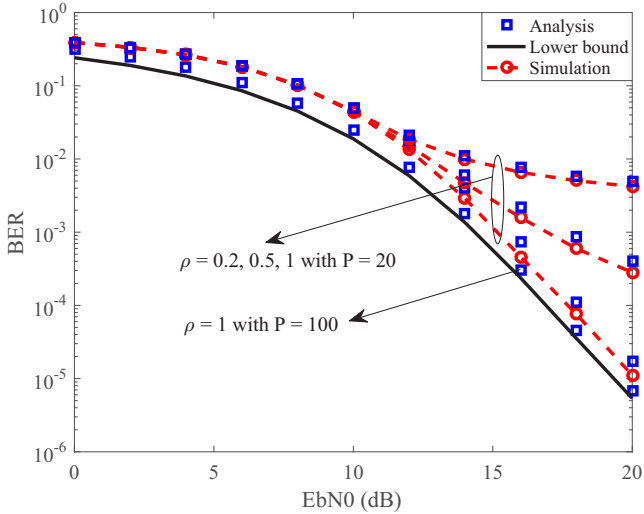}
  \caption{The system BERs under the PTJ and the BBJ environments with $JSR = \frac{P_j}{P_s} = 5$ dB.}
\end{figure}
\begin{figure}[!t]
  \includegraphics[width = 8.5cm]{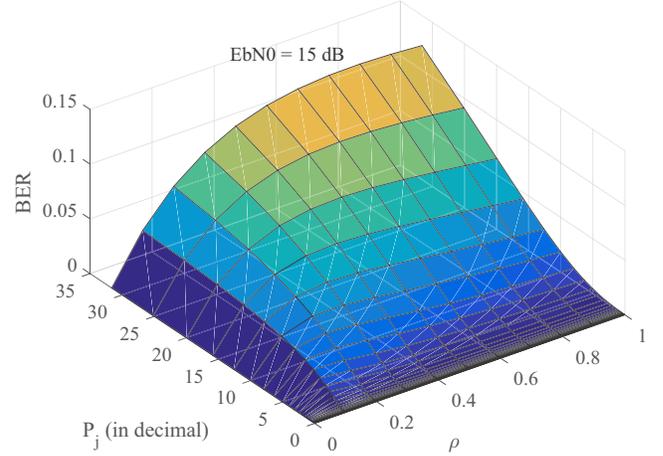}
  \caption{The system BERs versus $\rho$ and $P_j$ with $P = 20$ and $EbN0 = 15$ dB.}
\end{figure}

To further illustrate the effects of the jamming factor $\rho$ of the PTJ on the system performance, we present the system BER versus $\rho$ and $P_j$ in Fig. 5. We observe the trend that as $P_j$ is small, the optimal value of $\rho$ (from the jammer point of view) is close to zero. In addition, as $P_j$ increases, the optimal value of $\rho$ also increases and approaches one. This observation is understandable because when the jamming power is small, if the jamming is active all the time, the jamming power density will be very small, and thus, the jamming effect on the system performance could be negligible. Therefore, it is optimal for the jammer to turn on for a short period of time and transmit with a much higher power. On the other hand, when the jamming power is larger than a certain value, the composite jamming and AWGN power is comparable with (or even larger than) that of the targeted signal, and thus, it is unnecessary to further increase the jamming power. As a result, the jammer should turn on all the time to optimally degrade the target system performance. It is noteworthy that given $E_b/N_0$, $\beta$, $P$, and $P_j$, the optimal value of $\rho$ can be easily derived by numerically solving the optimization problem, whose objective function is the equation (20), with mathematical programs such as Matlab or Mathematica, i.e. with $E_b/N_0 = 15$ dB, $\beta = 200$, $P = 20$, and $P_j = 10$ dB, the optimal value of $\rho$ is found to be $0.24576$.
\begin{figure}[!t]
  \includegraphics[width = 8.5cm]{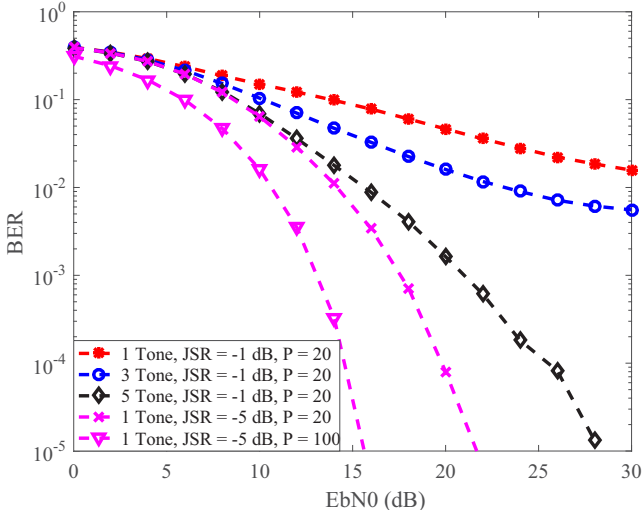}
  \caption{The system BERs versus $EbN0$ under the TJ environment.}
\end{figure}

We now turn our interest to the effect of the TJ on the performance of the NR-DCSK system. In the simulation, we use the following discrete baseband model of the TJ signals \cite{Lau-02}
\begin{align}
{j_k} = \sum\limits_{m = 1}^M {\sqrt {\frac{{2{P_j}}}{M}} \sin \left( {\pi k\frac{{F_m}}{{\beta }}} + \theta_m \right)},
\end{align}
where $F_m =f_m T_b$ is the normalized jamming frequency of the $m^{th}$ tone and $T_b$ denotes the bit duration. Here, $\theta_m$ is modeled as an arbitrary constant angle selected from $\left[ { - \pi ,\pi } \right]$. In addition, we simulate the system BER under single-tone, 3-tone, and 5-tone cases. It is shown in Fig. 6 that the STJ causes the most significant performance degradation and increasing the number of jamming tones actually enhances the performance of the target system. Moreover, for the STJ case, increasing $P$ can improve the system performance. The reason is that with a larger value of $P$, the receiver can further reduces the composite power of jamming and AWGN (by using the BAF), and thus, the SINR at the receiver is improved, from which the system BER is enhanced. Finally, it is observed that as $JSR$ is reduced, the system BER decreases, as expected.
\begin{figure}[!t]
  \includegraphics[width = 8.5cm]{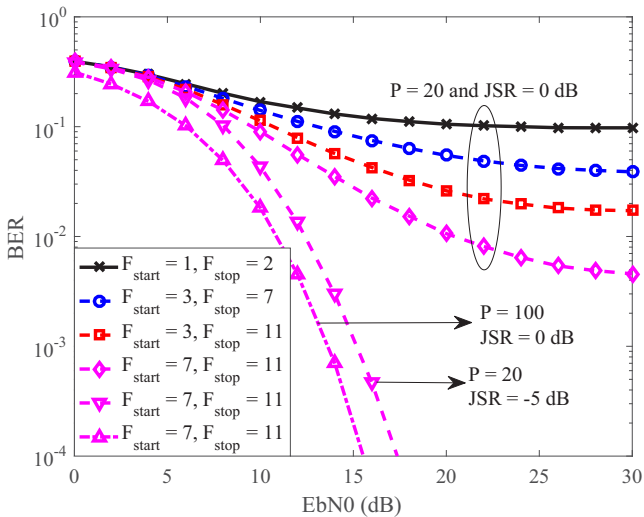}
  \caption{The system BERs versus $EbN0$ under the SWJ environment.}
\end{figure}
\begin{figure}[!t]
  \includegraphics[width = 8.5cm]{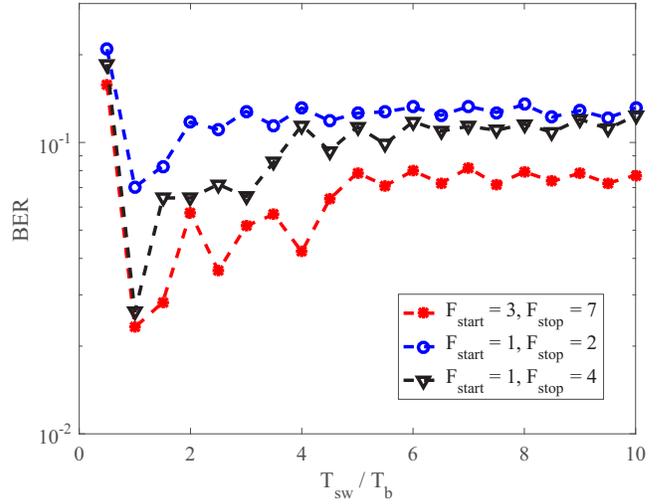}
  \caption{The system BERs versus $T_{sw}$ under the SWJ environment with $P = 20$ and $JSR = 0$ dB.}
\end{figure}

The effects of the SWJ on the performance of the NR-DCSK system will now be presented. According \cite{Glisic-95}, we can use the following discrete baseband model of the SWJ signals in our simulation
\begin{align}
j\left( k \right) = \sqrt {2{P_j}} \sin \left( {\pi \frac{{k{F_{start}}}}{\beta } + \pi \frac{{{k^2}\Delta F}}{{4{\beta ^2}}} + {\theta _{sw}}} \right),
\end{align}
where $F_{start} = f_{start} T_b$ and $\Delta F = \Delta f T_b^2$. In Fig. 7, we simulate the system BER versus $EbN0$ with the condition that the sweep time equals to the bit time, from which $\Delta F = {F_{stop}} - {F_{start}}$, where $F_{stop} = f_{stop} T_b$. The figure shows that the sweep frequencies and the sweep bandwidth have a great impact on the system performance. Particularly, the closer to the carrier frequency of the chaotic signals the starting frequency of the sweep jammer is and the smaller the sweep bandwidth is, the more significant the system performance is degraded. In addition, as expected, with a larger $P$ or with a smaller $JSR$, the system performance is enhanced.

In Fig. 8, we present the system BER versus the ratio of the sweep time and the bit duration with $P = 20$ and $JSR = 0$ dB. It can be seen that the most effective sweep time is half of the bit duration. The reason is that when $T_{sw} = T_b/2$, the reference and the information bearing sequences of any bit suffer from the same SWJ signals, which increases the correlation between the two sequences regardless of the bit value. Consequently, the decoding performance is degraded. In addition, the figure shows that when the sweep time is much larger than the bit duration, its effect on the system performance is negligible.

\section{Conclusion}
In this work, we considered the NR-DCSK system under the BBJ, PTJ, TJ, and the SWJ environments. We analytically derived closed-form expressions of the system BERs under the BBJ and the PTJ environments, from which we revealed that in general increasing $P$ can enhance the system anti-jamming performance. In addition, we showed that the optimal value of the jamming factor of the PTJ tends to zero when the jamming power is small, however, it increases and approaches one as the jamming power increases. Moreover, via simulations, we demonstrated that the STJ is more efficient than the MTJ counterparts, from the tone jammers' point of view. Furthermore, we illustrated that under the SWJ environment, the system performance is more significantly degraded when the starting frequency of the sweep jammer is closer to the carrier frequency of the chaotic signals and the sweep bandwidth is small. Finally, we revealed that the most effective sweep time is half of the bit duration. Our results could be use as a guideline for systems' designers in turning existing systems' parameters to satisfy a certain quality-of-service or designing novel and effective AJ communication systems.

\section*{Acknowledgment}
The authors gratefully acknowledge the support from Electronic Warfare Research Center at Gwangju Institute of Science and Technology (GIST), originally funded by Defense Acquisition Program Administration (DAPA) and Agency for Defense Development (ADD).

\ifCLASSOPTIONcaptionsoff
  \newpage
\fi

\end{document}